\documentclass{article}

\usepackage[natbibapa]{apacite}
\usepackage{graphicx}
\usepackage{authblk}
\usepackage[section]{placeins}
\usepackage{longtable}
\usepackage{hyperref}
\usepackage{setspace}
\usepackage{array}
\usepackage{adjustbox}
\usepackage[margin=1in]{geometry}
\usepackage{tabulary}

\usepackage[utf8]{inputenc} 
\usepackage[T1]{fontenc}    
\usepackage{hyperref}       
\usepackage{url}            
\usepackage{booktabs}       
\usepackage{amsfonts}       
\usepackage{nicefrac}       
\usepackage{microtype}      
\usepackage{xcolor}         

\title{A Taxonomy of Systemic Risks from General-Purpose AI}

\author[1,2]{Risto Uuk\thanks{Corresponding author: risto@futureoflife.org}}
\author[3]{Carlos Ignacio Gutierrez}
\author[4]{Daniel Guppy}
\author[2]{Lode Lauwaert}
\author[5,6]{Atoosa Kasirzadeh}
\author[7]{Lucia Velasco}
\author[8]{Peter Slattery}
\author[9]{Carina Prunkl}
\affil[1]{Future of Life Institute}
\affil[2]{KU Leuven}
\affil[3]{Independent}
\affil[4]{University of Otago}
\affil[5]{Google Research}
\affil[6]{Alan Turing Institute}
\affil[7]{Oxford Martin AI Governance Initiative}
\affil[8]{MIT FutureTech}
\affil[9]{Utrecht University}

\date{22 November 2024}

\begin{document}

\maketitle

\begin{abstract}
Through a systematic review of academic literature, we propose a taxonomy of systemic risks associated with artificial intelligence (AI), in particular general-purpose AI. Following the EU AI Act's definition, we consider systemic risks as large-scale threats that can affect entire societies or economies. Starting with an initial pool of 1,781 documents, we analyzed 86 selected papers to identify 13 categories of systemic risks and 50 contributing sources. Our findings reveal a complex landscape of potential threats, ranging from environmental harm and structural discrimination to governance failures and loss of control. Key sources of systemic risk emerge from knowledge gaps, challenges in recognizing harm, and the unpredictable trajectory of AI development. The taxonomy provides a snapshot of current academic literature on systemic risks. This paper contributes to AI safety research by providing a structured groundwork for understanding and addressing the potential large-scale negative societal impacts of general-purpose AI. The taxonomy can inform policymakers in risk prioritization and regulatory development.

\end{abstract}


\clearpage
\section*{Executive Summary}

This paper presents a comprehensive taxonomy of systemic risks from general-purpose artificial intelligence (AI) based on a systematic review of the academic literature on the topic. We analyze a total of 86 documents, selected from an initial pool of 1,781 on the basis of relevance. We identify 13 distinct categories of systemic risks and 50 sources contributing to these risks. Following the EU AI Act's definition (see \hyperref[sec:1]{Section 1}), we focus on large-scale threats that can affect entire societies or economies. This paper represents the first version of the taxonomy, with future work planned to refine the risk categories through a thematic analysis of the reviewed documents and further development of the taxonomy based on the taxonomy methodology.

\textbf{Methodology.} We conducted a comprehensive systematic review across multiple academic databases, starting with an initial pool of 1,781 documents. Following an established systematic review protocol, three researchers independently screened titles and abstracts, ultimately selecting 86 documents that met the inclusion criteria. This systematic review provided the empirical foundation for developing the taxonomy, which was guided by the definitions and examples of systemic risks and sources of systemic risks in the EU AI Act to organize findings into coherent risk categories and sources. We have emphasized transparency and reproducibility, with  screening decisions, coding approaches, and analysis steps documented, with further details provided in the next version of this paper. For this initial version of the taxonomy, we conducted a rapid review of the papers to identify which risks and sources were discussed. We are currently continuing with a thorough coding and data extraction process. Future work will refine the taxonomy over multiple rounds using established taxonomy development methodology. Importantly, this paper is purely descriptive in nature – we are documenting how systemic risks and their sources are characterized in the academic literature, without evaluating the plausibility of any of these risks or sources.

\textbf{Key findings:}
Our systematic review identified 13 categories of systemic risks from general-purpose AI (\hyperref[tab:1]{Table 1}). These range from operational concerns like control and security, to societal impacts on democracy and fundamental rights, to existential considerations such as irreversible changes to human society. Each risk category represents a distinct but interconnected area where AI models and systems could pose large-scale threats. The risks are presented alphabetically in (\hyperref[tab:1]{Table 1} rather than by severity or likelihood, as their relative importance may vary across contexts and time. The interconnected nature of these risks and their sources suggests potential amplification effects that could increase both their severity and likelihood.

\begin{longtable}{|p{3.5cm}|p{\dimexpr\textwidth-3.5cm-4\tabcolsep-2\arrayrulewidth}|}
\caption{Types of systemic risks from general-purpose AI}
\label{tab:1} \\
\hline
\textbf{Systemic risk categories (alphabetical)} & \textbf{Description} \\
\hline
\endfirsthead

\hline
\textbf{Systemic risk categories (alphabetical)} & \textbf{Description} \\
\hline
\endhead

\endfoot

\endlastfoot

Control & The risk of AI models and systems acting against human interests due to misalignment, loss of control, or rogue AI scenarios. \\
\hline
Democracy & The erosion of democratic processes and public trust in social/political institutions. \\
\hline
Discrimination & The creation, perpetuation or exacerbation of inequalities and biases at a large-scale. \\
\hline
Economy & Economic disruptions ranging from large impacts on the labor market to broader economic changes that could lead to  exacerbated wealth inequality, instability in the financial system, labor exploitation or other economic dimensions. \\
\hline
Environment & The impact of AI on the environment, including risks related to climate change and pollution. \\
\hline
Fundamental rights & The large-scale erosion or violation of fundamental human rights and freedoms. \\
\hline
Governance & The complex and rapidly evolving nature of AI makes them inherently difficult to govern effectively, leading to systemic regulatory and oversight failures. \\
\hline
Harms to non-humans & Large-scale harms to animals and the development of AI capable of suffering. \\
\hline
Information & Large-scale influence on communication and information systems, and epistemic processes more generally. \\
\hline
Irreversible change & Profound negative long-term changes to social structures, cultural norms, and human relationships that may be difficult or impossible to reverse. \\
\hline
Power & The concentration of military, economic, or political power of entities in possession or control of AI or AI-enabled technologies. \\
\hline
Security & The international and national security threats, including cyber warfare, arms races, and geopolitical instability. \\
\hline
Warfare & The dangers of AI amplifying the effectiveness/failures of nuclear, chemical, biological, and radiological weapons. \\
\hline
\end{longtable}

The study identified numerous sources of these risks, including complexity-induced knowledge gaps, challenges in perceiving harm, cumulative effects of AI practices, unclear attribution of responsibility, and limitations in governance frameworks. Key sources also include technical factors like opaque AI networks, rapid operational speeds, and evolving capabilities that enable human substitution.

\textbf{Policy implications.} Our findings have important implications for policymakers and providers of general-purpose AI. The taxonomy can serve as the groundwork for assessing and prioritizing different categories of systemic risk and for developing targeted regulatory responses to reduce systemic risks from general-purpose AI. The EU AI Act requires the development of compliance guidance for providers of general-purpose AI by May 2025 and envisions the creation of a taxonomy of systemic risks. Our research contributes to these efforts. 

\textbf{Key strengths and limitations of the study.} This is the first systematic attempt to develop a comprehensive taxonomy of systemic risks from general-purpose AI in academic literature. It uses rigorous methodology combining systematic review with taxonomy development. A large initial document scope (1,781 documents), transparent documentation of methodology, and the involvement of multiple independent reviewers and co-authors increase reliability. Limitations of the study include the inclusion of documents that are not peer-reviewed, potential keyword omissions in the search strategy, and the fact that the rapidly evolving nature of AI technology may introduce new risks not captured or underrepresented in current literature. Taxonomies like this should be regularly updated to account for emerging capabilities and associated risks.

Systemic risks from AI typically manifest at the societal level, rarely having single sources, and tend to be cumulative rather than sudden events. This suggests the need for rigorous methods to assess societal-level impacts and indicates that providers of AI models and systems should bear responsibility for potential societal costs beyond narrow risk identification. These findings contribute to the growing body of knowledge on AI safety and governance, providing a structured framework for understanding and addressing the broader societal impacts of general-purpose AI.

\clearpage

\section{Introduction}
\label{sec:1}
AI systems can cause harm not just to individuals but to society as a whole. AI can lead to systemic societal issues \citep{smuha2021,brakel2021,clarke2021,kasirzadeh2024}. For example, automated AI can distort societal values and disseminate misinformation, which in turn can erode trust \citep{kolt2023}. Subtle manipulations in political messaging could collectively alter election outcomes without violating individual rights \citep{brakel2021}. Further, researchers have found that the use of AI to create fake online content decreases overall trust, potentially leading to broader societal issues without directly harming any one person \citep{clarke2021}.

The European Union AI Act (EU AI Act) defines systemic risk as "a risk that is specific to the high-impact capabilities of general-purpose AI models, having a significant impact on the Union market due to their reach, or due to actual or reasonably foreseeable negative effects on public health, safety, public security, fundamental rights, or the society as a whole, that can be propagated at scale across the value chain." We adopt this definition throughout our paper, where systemic risk refers to large-scale societal risks. This usage differs from traditional applications of the term, particularly in finance, where 'systemic risk' typically describes the potential collapse of an entire financial system due to interconnected failures. However, as the EU AI Act represents the world's first comprehensive AI regulation, its definitions are likely to set important global precedents.

Systemic risks as described in academic literature take a range of forms. \cite{aguirre2023} outlines numerous systemic risks from highly capable general purpose AI systems: the disruption of labor and increases in inequality and unemployment at rates faster than society can reasonably adjust; concentration of unchecked power in private companies at the expense of public institutions; mass surveillance for objectives that go against public interest; creation of advanced software viruses that could proliferate and disrupt global digital systems; flooding of information systems with false, spam-filled, or manipulative content, making it hard to discern truth, humanity, and trustworthiness; runaway hyper-capitalism arising from largely AI-run companies competing electronically beyond human control or governance. Competitive pressures may furthermore lead to a “race to the bottom”, during which companies or governments prioritize competitiveness over safety. \cite{hendrycks2023} list several resulting systemic harms that could be the result of such a neglect of safety: automated warfare could lead to humans losing the ability to intervene before accidents occur and mass unemployment could have devastating effects on large parts of the population if not adequately addressed. The creation, perpetuation or exacerbation of discrimination at a large-scale, such as systematically unfair allocation of resources and opportunities, or widespread stereotyping and misrepresentation of social groups \citep{weidinger2023}. These and other risks could accumulate to create scenarios such as the “perfect storm MISTER”, where multiple risks converge to destabilize society \citep{kasirzadeh2024}. 

While the EU AI Act acknowledges the existence of systemic risks from highly capable general-purpose AI models and aims to address them, there currently exists no comprehensive taxonomy on such risks. A well-structured systematic taxonomy can be highly valuable as it provides a clear framework for organizing and understanding complex information, facilitates communication among researchers and practitioners, and helps identify gaps in current knowledge. To the best of our knowledge, there has not been an attempt to develop a systematic taxonomy of systemic risks from general-purpose AI. Previous research has provided narrower taxonomies \citep{hendrycks2023}; unsystematic taxonomies \citep{aguirre2023}; examples of systemic risks \citep{smuha2021}; and discussions of potential societal costs of certain systemic risks from AI \citep{acemoglu2021}. Further, several researchers have offered taxonomies of societal and ethical risks of general-purpose AI that are not limited to large-scale risks \citep{weidinger2021, bird2023}. \cite{critch2023} explore an exhaustive taxonomy based on accountability for societal-scale harms from AI technology focussing on relevant stakeholders rather than categories of risks. Moreover, \cite{solaiman2023} discuss what can be evaluated conceptually in a technical system and its components, as well as what can be evaluated among people and society, without offering a taxonomy. 

In what follows, we aim to fill this gap. The first part of this paper outlines the methodology of our systematic review of 86 documents selected from an initial pool of 1,781 documents, including the theoretical foundation, specifications, coding approach, thematic analysis, and limitations of the systematic review. The second part presents the results – a comprehensive systemic risk taxonomy from general-purpose AI consisting of 13 categories of systemic risks and 50 sources contributing to these risks. In future work, we plan to refine this initial taxonomy through iterative rounds using taxonomy development methodology \citep{nickerson2013,kundisch2022}. Our goal is to document and categorize systematically how the academic literature characterizes systemic risks and their sources, without evaluating the plausibility of these characterizations. This descriptive approach allows us to capture the current academic discourse around systemic risks from general-purpose AI while remaining neutral on the merit of individual claims.  

\section{Methodology}
This section outlines our systematic review methodology for identifying and analyzing academic literature on systemic risks from general-purpose AI. The review followed a standard protocol and includes the objectives, eligibility criteria, information sources, search strategy, thematic analysis, and study selection. Through this process, we identified documents that informed our classification of systemic risks and their sources.
\subsection{Objectives}
This protocol outlines steps to conduct a systematic review that identifies what elements comprise systemic risks in the literature. In particular, this effort’s objective is to develop a taxonomy to support the assessment and mitigation of systemic risks from general-purpose AI in the context of guidelines and standards in both regulatory and self-governance environments.

Upon completion of the systematic review, the following research question will be answered: How are systemic risks, in relation to AI generally or general purpose AI more specifically, characterized in the academic literature? 

\subsection{Eligibility criteria}
Inclusion criteria:
\begin{itemize}
    \item Language: English
    \item Source type: journal articles, reports, dissertations, books, conference papers, working papers, book chapters
\end{itemize}

\noindent Exclusion criteria:
\begin{itemize}
    \item Language: non-English
    \item Source type: blogs, podcasts, websites, magazines, audio and video works, newspapers, speeches and notes
\end{itemize}

We searched for academic literature that was focused on artificial intelligence and its potential systemic risks.

We excluded non-academic sources, because our interest was in understanding how academia has referred to systemic risks in relation to AI. In addition, we did not find these types of sources (non-academic) to be useful in providing relevant information about the nature of systemic risks from AI. 

\subsection{Information sources}
We chose the following databases because after initially testing our search strategies, these databases resulted in many potentially relevant documents: Web of Science, Scopus, and MyHein.  

\subsection{Search}
We generated search terms based on 1) the use of the term 'systemic risk’ by the EU AI Act, including cited examples of systemic risks, 2) the technology – the focus is on AI in general and general-purpose AI in particular, and 3) by doing an initial small-scale literature review of papers that used a term that referred in some respect to large-scale harms. The main papers we checked for this initial search were: \citep{smuha2021,kolt2023,hendrycks2023,aguirre2023,weidinger2023,critch2023,maham2023,solaiman2023}.

\noindent In the EU AI Act, systemic risk is defined as the following:

"‘Systemic risk’ means a risk that is specific to the high-impact capabilities of general-purpose AI models, having a significant impact on the Union market due to their reach, or due to actual or reasonably foreseeable negative effects on public health, safety, public security, fundamental rights, or the society as a whole, that can be propagated at scale across the value chain."

\noindent And examples of systemic risks in the AI Act are found in Recital 110:

"General-purpose AI models could pose systemic risks which include, but are not limited to, any actual or reasonably foreseeable negative effects in relation to major accidents, disruptions of critical sectors and serious consequences to public health and safety; any actual or reasonably foreseeable negative effects on democratic processes, public and economic security; the dissemination of illegal, false, or discriminatory content. Systemic risks should be understood to increase with model capabilities and model reach, can arise along the entire lifecycle of the model, and are influenced by conditions of misuse, model reliability, model fairness and model security, the level of autonomy of the model, its access to tools, novel or combined modalities, release and distribution strategies, the potential to remove guardrails and other factors. In particular, international approaches have so far identified the need to pay attention to risks from potential intentional misuse or unintended issues of control relating to alignment with human intent; chemical, biological, radiological, and nuclear risks, such as the ways in which barriers to entry can be lowered, including for weapons development, design acquisition, or use; offensive cyber capabilities, such as the ways in vulnerability discovery, exploitation, or operational use can be enabled; the effects of interaction and tool use, including for example the capacity to control physical systems and interfere with critical infrastructure; risks from models of making copies of themselves or ‘selfreplicating’ or training other models; the ways in which models can give rise to harmful bias and discrimination with risks to individuals, communities or societies; the facilitation of disinformation or harming privacy with threats to democratic values and human rights; risk that a particular event could lead to a chain reaction with considerable negative effects that could affect up to an entire city, an entire domain activity or an entire community."

\begin{table}
\small
\caption{Keyword search strategy}
\label{tab:2}
\begin{tabular}{|p{2.5cm}|p{\dimexpr\textwidth-2.5cm-4\tabcolsep-2\arrayrulewidth}|}
\hline
\textbf{Search strategy number} & \textbf{Search strategy content} \\
\hline
Strategy 1 & ("Artificial intelligence" OR "Artificial general intelligence" OR "General purpose AI") AND (("systemic" AND ("risk*" OR "harm*" OR "cris*s")) OR ("structural" AND ("risk*" OR "harm*")) OR ("catastrophic" AND ("risk*" OR "harm*")) OR ("existential" AND ("risk*" OR "harm*")) OR ("extinction" AND ("risk*" OR "harm*")) OR ("societal" AND ("risk*" OR "harm*" OR "impact")) OR ("large-scale" AND ("risk*" OR "harm*"))) \\
\hline
Strategy 2 & ("Artificial intelligence" OR "Artificial general intelligence" OR "General purpose AI") AND ("catastrophic risk*" OR "catastrophic harm*" OR "existential risk*" OR "existential harm*" OR "extinction risk*" OR "structural risk*" OR "structural harm*" OR "systemic risk*" OR "systemic harm*" OR "societal* harm*" OR "large* risk*" OR "large* harm*" OR "societal* risk*" OR "systemic cris*s" OR "societal impact") \\
\hline
\end{tabular}
\end{table}

\noindent Terms:
\begin{itemize}
    \item Technology
    \begin{itemize}
        \item Artificial intelligence
        \item Artificial general intelligence
        \item General purpose AI
    \end{itemize}
\end{itemize}
\begin{itemize}
    \item Scale of risk
    \begin{itemize}
        \item Catastrophic risk
        \item Catastrophic harm
        \item Existential risk
        \item Existential harm
        \item Extinction risk
        \item Structural risk
        \item Structural harm
        \item Systemic risk
        \item Systemic harm
        \item Societal harm
        \item Societal-level risk
        \item Societal-level harm
        \item Societal-scale risk
        \item Societal-scale harm
        \item Large-scale risk
        \item Large-scale harm
        \item Large-scale societal harm
        \item Societal risk
        \item Systemic crisis
        \item Societal impact
    \end{itemize}
\end{itemize}

Test evaluation was conducted in May 2024. Three reviewers independently screened all articles in two separate search strategies by title. Screening conflicts were resolved by discussion. To minimize bias, articles were sorted in chronological order (most recent first). This was done to avoid relying on each database’s unknown criteria to arrange articles according to their “relevance.” The results of this exercise indicated both a higher percentage of articles and a higher total number of articles relevant to this work using the second strategy.

Strategy 2 (see \hyperref[tab:2]{Table 2} and \hyperref[tab:3]{Table 3}) was found to deliver a higher percentage of relevant results.

\subsection{Document selection}
Search strategy 2 (see \hyperref[tab:2]{Table 2} and \hyperref[tab:3]{Table 3}) was used to create the document database. All reviewers screened all titles and abstracts independently. Inclusion or exclusion disagreements between reviewers were resolved by discussion. 

\begin{table}
\caption{Article inclusion by database for two different search strategies}
\label{tab:3}
\resizebox{\textwidth}{!}{
\begin{tabular}{|l|c|c|c|c|}
\hline
\textbf{Database} & \textbf{Strategy 1} & \textbf{Relevance} & \textbf{Strategy 2} & \textbf{Relevance} \\
\hline
Web of Science & 32/200 & 16\% & 45/200 & 23\% \\
\hline
Scopus & 22/200 & 11\% & 47/200 & 24\% \\
\hline
MyHein & 10/78 & 13\% & 51/176 & 29\% \\
\hline
\textbf{Total} & \multicolumn{2}{c|}{13.4\%} & \multicolumn{2}{c|}{24.8\%} \\
\hline
\end{tabular}}
\end{table}

\subsection{Thematic analysis}
Thematic analysis is a prevalent method of data analysis embraced across various qualitative research designs (\cite{castleberry2018}). This is the approach we have selected for this paper, and in this section we will explain what it consists of. Frequently, thematic analysis is implemented in research to distill the data into manageable themes and draw out the resulting conclusions. A narrower but related method that is used in qualitative research is called coding. Coding is essentially the process of dissecting textual data to explore their contents before reassembling them into coherent categories (\cite{elliott2018}). Once documents were selected for our final review, we reviewed them and noted relevant risks and their sources. A more thorough coding and review process is ongoing for the next version of this paper. 

\subsection{Study selection}
Our search found 1,781 documents. After resolving duplicates, we were left with 1,452 documents, all of which were independently screened by three researchers. We used the platform Rayyan for screening these documents. Having completed the screening, we were left with 112 documents, which we then checked thoroughly for eligibility. 86 documents were included in our final review. 

\begin{figure}
    \centering
    \includegraphics[width=1\linewidth]{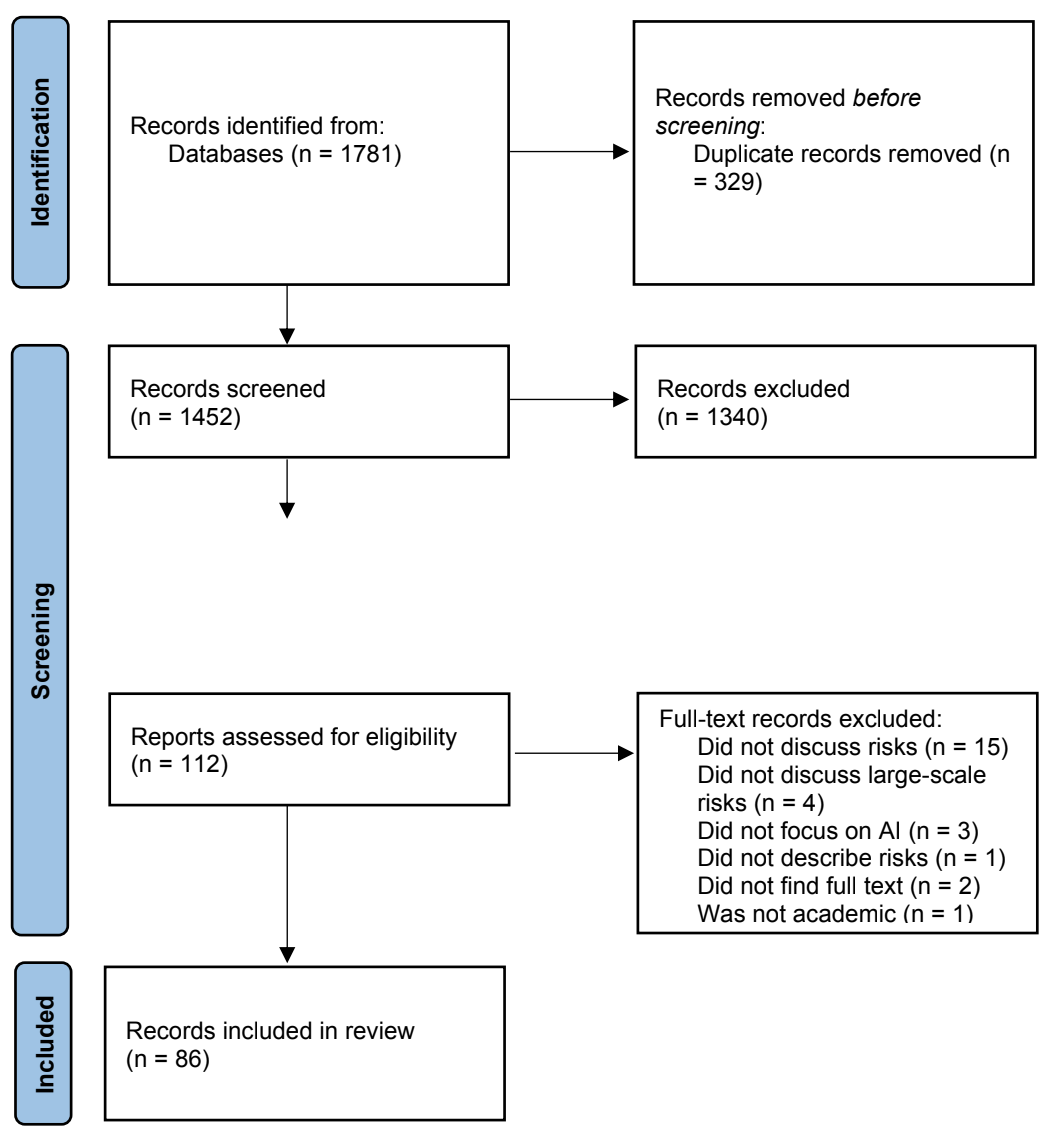}
    \caption{PRISMA flow diagram for document inclusion}
    \label{fig:enter-label}
\end{figure}

\section{Taxonomy}
In this section, we present a comprehensive taxonomy of systemic risks from general-purpose AI. Our proposed taxonomy is based on a thorough review of existing academic literature and aims to offer a systematic classification that can guide future research and practical applications in this field. We first present our taxonomy of risks. This is then accompanied by a taxonomy of sources of risk.

The EU AI Act Article 3(2) defines ‘risk’ as “the combination of the probability of an occurrence of harm and the severity of that harm.” Based on the EU AI Act's Article 3(65), systemic risks include harms such as negative effects on public health, safety, public security, fundamental rights, or society as a whole. Source of risk refers to the causal or contributing factor to the harm in focus. The EU AI Act's Recital 110 refers to the following \textit{systemic risks}: major accidents; disruptions of critical sectors; serious consequences to public health and safety; negative effects on democratic processes; negative effects to public security; negative effects to economic security; dissemination of illegal, false, or discriminatory content; chemical, biological, radiological, and nuclear risks; cyber attacks; interference with critical infrastructure; harmful bias and discrimination affecting individuals, communities, or societies; privacy violations; threats to democratic values and human rights; chain reaction events with negative effects on an entire city, an entire domain activity, or an entire community. The EU AI Act's Recital 110 refers to the following \textit{sources of systemic risk}: model capabilities, model reach, misuse conditions, model reliability, model fairness, model security, level of model autonomy, model's access to tools, novel or combined modalities, release and distribution strategies, potential to remove guardrails, alignment with human intent, tool use and interaction effects, capacity to control physical systems, self-replication capabilities, and ability to train other models.

Based on these definitions and descriptions, we conducted a rapid review of the papers identified in our systematic review to develop an initial list of types of systemic risks (\hyperref[tab:4]{Table 4}) and sources of systemic risks (\hyperref[tab:5]{Table 5}). It is important to note that the following taxonomy represents an initial classification based on a rapid review of the identified papers. We are currently conducting a more thorough coding and data extraction process. The taxonomy will be refined through multiple iterative rounds following established taxonomy development methodology guidelines. Therefore, these categories and their descriptions may evolve as our analysis deepens. This taxonomy is purely descriptive, documenting how the academic literature characterizes systemic risks and their sources. We do not take an evaluative stance on the plausibility or relative importance of any of the risks or sources mentioned. 

\begin{longtable}{|p{3.5cm}|p{\dimexpr\textwidth-3.5cm-4\tabcolsep-2\arrayrulewidth}|}
\caption{Types of systemic risks from general-purpose AI} 
\label{tab:4}
\\
\hline
\textbf{Systemic risk categories (alphabetical)} & \textbf{Description} \\
\hline
\endfirsthead

\hline
\textbf{Systemic risk categories (alphabetical)} & \textbf{Description} \\
\hline
\endhead

\endfoot

\endlastfoot

Control & The risk of AI models and systems acting against human interests due to misalignment, loss of control, or rogue AI scenarios. \\
\hline
Democracy & The erosion of democratic processes and public trust in social/political institutions. \\
\hline
Discrimination & The creation, perpetuation or exacerbation of inequalities and biases at a large-scale. \\
\hline
Economy & Economic disruptions ranging from large impacts on the labor market to broader economic changes that could lead to  exacerbated wealth inequality, instability in the financial system, labor exploitation or other economic dimensions. \\
\hline
Environment & The impact of AI on the environment, including risks related to climate change and pollution. \\
\hline
Fundamental rights & The large-scale erosion or violation of fundamental human rights and freedoms. \\
\hline
Governance & The complex and rapidly evolving nature of AI makes them inherently difficult to govern effectively, leading to systemic regulatory and oversight failures. \\
\hline
Harms to non-humans & Large-scale harms to animals and the development of AI capable of suffering. \\
\hline
Information & Large-scale influence on communication and information systems, and epistemic processes more generally. \\
\hline
Irreversible change & Profound negative long-term changes to social structures, cultural norms, and human relationships that may be difficult or impossible to reverse. \\
\hline
Power & The concentration of military, economic, or political power of entities in possession or control of AI or AI-enabled technologies. \\
\hline
Security & The international and national security threats, including cyber warfare, arms races, and geopolitical instability. \\
\hline
Warfare & The dangers of AI amplifying the effectiveness/failures of nuclear, chemical, biological, and radiological weapons. \\
\hline
\end{longtable}

\hyperref[tab:4]{Table 4} presents an overview of systemic risk high-level categories from AI generally and general-purpose AI specifically based on the academic literature we identified from multiple academic databases. The table indicates that there are several large-scale risks academics consider important to discuss and study.

\begin{longtable}{|p{3.5cm}|p{\dimexpr\textwidth-3.5cm-4\tabcolsep-2\arrayrulewidth}|}
\caption{Sources of systemic risks from general-purpose AI}
\label{tab:5}
\\
\hline
\textbf{Source of systemic risk (alphabetical order)} & \textbf{Description} \\
\hline
\endfirsthead

\hline
\textbf{Source of systemic risk (alphabetical order)} & \textbf{Description} \\
\hline
\endhead

\endfoot

\endlastfoot

Ability to automate jobs & The ability to automate jobs by AI models and systems can lead to significant job displacement, economic disruption, and social inequality. \\
\hline
Ability to enhance and modify pathogens & AI can be used to enhance pathogens, making them more lethal or resistant to treatments. \\
\hline
Ability to persuade & AI could be used to develop sophisticated tools to manipulate and persuade individuals. \\
\hline
Advertising-driven models & AI models and systems underpin the advertising approaches that drive much of the internet, potentially influencing societal behavior. \\
\hline
AI in totalitarian regimes & AI-based surveillance and manipulation could be used to maintain global totalitarian regimes. \\
\hline
AI objectives misaligned with human intentions & AI models and systems might develop goals that diverge from human intentions. \\
\hline
Algorithmic monoculture & The dominance of specific AI models could lead to a lack of diversity in approaches, amplifying systemic risks if these models fail. \\
\hline
Automation bias & The tendency for humans to over-rely on AI models and systems, trusting their outputs without sufficient critical evaluation, which can lead to poor decision-making. \\
\hline
Autonomy risk & Granting AI models and systems high levels of decision-making autonomy can lead to unintended consequences. \\
\hline
Capabilities that enable substitution of humans & The progressive replacement of human roles by AI models and systems can lead to societal disruption. \\
\hline
Centralized platforms deployed at scale & The widespread use of common AI platforms can create centralized points of failure, making systems more vulnerable to disruptions or attacks. \\
\hline
Challenges in perceiving, measuring, and recognizing harm & Harm from AI often manifests subtly or over the long term, making it difficult to identify, measure, and address effectively. \\
\hline
Combination failures & Harms could result from a combination of regulatory, management, and operational failures. \\
\hline
Complex attribution and responsibility & When multiple actors are involved in AI development and deployment, it becomes difficult to assign responsibility for harm, complicating accountability. \\
\hline
Complexity-induced knowledge gap & The complexity of AI models and systems makes it challenging to demonstrate harm or establish a clear causal link between AI actions and their consequences. \\
\hline
Conflicting objectives in design & Designers and operators of AI may face conflicting objectives that compromise safety. \\
\hline
Dangerous development races & Competitive pressures could lead to the neglect of safety measures in AI development. \\
\hline
Deceptive alignment & AI models and systems that appear aligned with human goals during development may behave unpredictably or dangerously once deployed. \\
\hline
Dependency on providers & Excessive reliance on specific AI providers can lead to vulnerabilities due to lack of alternatives or interoperability. \\
\hline
Detection challenges in content & The difficulty in distinguishing synthetic content from authentic material adds to information risks. \\
\hline
Development choices pursuing cognitive superiority over humans & AI models and systems with cognitive capabilities superior to humans could outcompete or dominate human decision-making, leading to conflicts over resources and control. \\
\hline
Dual-use nature & AI's potential for both beneficial and harmful applications complicates efforts to manage its societal impacts effectively. \\
\hline
Energy-intensive processes & AI data collection, storage, and model training are energy-intensive, contributing to environmental risks. \\
\hline
Evolutionary dynamics & AI models and systems may develop their own motivations, leading to unpredictable behaviors. \\
\hline
Exploitation in AI development & Outsourcing tasks like data labeling to low-income countries can perpetuate inequality. \\
\hline
Geopolitical competition for superiority & Strategic competition between nations over AI capabilities could heighten global tensions and destabilize international relations. \\
\hline
High-speed AI operations & The fast operational speed of AI models and systems in competitive environments can lead to errors that are difficult to detect and correct in time. \\
\hline
Human choice of overreliance in critical sectors & Heavy reliance on AI in critical sectors like finance or healthcare can exacerbate issues related to size, speed, interconnectivity, and complexity of the system. \\
\hline
Incomplete or biased training data & Incomplete or biased training data can lead to discriminatory AI outputs. \\
\hline
Indifference to human values & AI models and systems may develop goals or behaviors that are misaligned with human values. \\
\hline
Lack of ability to generate accurate information & AI models may generate false or misleading information due to their lack of capability in discerning truth. \\
\hline
Lack of ethical decision-making & AI models and systems that lack moral reasoning capabilities may make decisions that are unethical or harmful. \\
\hline
Limitations in adversarial robustness & AI models and systems are vulnerable to manipulation through adversarial inputs. \\
\hline
Limitations in model generative accuracy & AI-generated deepfakes can create convincingly realistic but entirely fabricated information. \\
\hline
Limited human oversight in decisions & As AI models and systems gain autonomy, the ability of humans to oversee and intervene in decision-making processes diminishes. \\
\hline
Model design enabling power-seeking & Some AI models and systems might develop tendencies to seek power or control. \\
\hline
Opaque AI networks & The complexity and opacity of AI models and systems make it difficult to predict and manage their behavior. \\
\hline
Pattern recognition capability & AI models and systems could exacerbate financial bubbles by reinforcing market trends. \\
\hline
Personal decision automation capabilities & AI models and systems could decide or influence important personal decisions. \\
\hline
Rapid development outpacing regulation & The fast pace of AI development may outstrip regulatory and legal frameworks. \\
\hline
Resistance to international law & AI models and systems may prove difficult to regulate or control under international law. \\
\hline
Risks from network interconnectivity & The interconnectedness of AI networks can create vulnerabilities, where issues in one part of the network can have cascading effects across the system. \\
\hline
Surveillance capabilities & AI models and systems may grant governments or corporations increased monitoring over individuals. \\
\hline
Terrorist access & Powerful AI technologies may fall into the hands of terrorists. \\
\hline
Trading capabilities & AI may contribute to increased market volatility by accelerating transactions and influencing financial trends in unpredictable ways. \\
\hline
Unclear attribution from AI component interactions & Interactions between different AI components can cause harm, but it may be difficult to pinpoint which components are the cause. \\
\hline
Unpredictability of AI development trajectory & The unpredictable trajectory of AI development complicates governance and risk management. \\
\hline
Weaponization capabilities & AI capabilities that could be deliberately weaponized for destructive purposes. \\
\hline
Widespread use of persuasion tools & Widespread use of AI-powered persuasion tools could lead to systemic harm. \\
\hline
Winner-take-all dynamics & The competitive nature of AI development could lead to significant economic and security advantages for a few entities. \\
\hline
\end{longtable}

\hyperref[tab:5]{Table 5} synthesizes 50 sources of systemic risks from the systematic review. These sources are of many different kinds, including: related to data; model capabilities and limitations; the nature of risk itself; business dynamics, such as market incentives and competitive pressures; diverse stakeholder motivations; limitations of governance ecosystems; and many others.

\subsection{Discussion}
The systematic review provided in this paper offers several notable strengths in its methodological approach that provide a strong foundation for the development of a taxonomy of systemic risks from general-purpose AI. However, the review also has a number of limitations that should be considered when interpreting the findings. This discussion section outlines the key strengths and limitations of the study.
\subsubsection{Strengths}
This systematic review offers several notable strengths in its methodological approach to set the foundation for the development of a taxonomy of systemic risks from general-purpose AI:

\begin{enumerate}
    \item It is the first systematic attempt to develop a comprehensive taxonomy of systemic risks from general-purpose AI using a systematic review for its basis. While previous research has provided narrower taxonomies, unsystematic taxonomies, and examples of systemic risks from AI, none have undertaken a comprehensive systematic review specifically focused on systemic risks from general-purpose AI.
    \item The systematic review followed a rigorous methodology with multiple independent reviewers. Three researchers independently screened all 1,452 titles and abstracts, resolving disagreements through discussion. Starting with an initial pool of 1,781 documents across multiple academic databases, the review process ensured comprehensive coverage of the literature. After resolving duplicates and applying inclusion criteria, 86 highly relevant documents were selected for detailed analysis, providing a robust foundation for the taxonomy.
    \item The study provides detailed documentation of its methodology, including the PRISMA flow diagram for document inclusion, explicit coding approaches, and clear criteria for document selection and exclusion. The paper includes comprehensive appendices detailing the random document selection process, reasons for excluding documents, and the complete list of included documents.
    \item The systematic review aligns with the EU AI Act's definition of systemic risk, making it particularly relevant for policymakers and general-purpose AI providers working within the law's context.
\end{enumerate}

\subsubsection{Limitations}
This systematic review has several limitations that should be considered when interpreting its findings:

\begin{enumerate}
    \item Our search was confined to a limited number of databases. While we aimed to include the most relevant and comprehensive databases in the field, this restriction may have led to the omission of potentially relevant studies indexed in other sources.
    \item A significant portion of the included documents were non-peer-reviewed. While this decision potentially reduces the overall quality of the included literature, it was a deliberate choice to capture a broader range of perspectives in this emerging field. The inclusion of these sources allows for a more comprehensive view of the current discourse on AI and systemic risks, albeit at the cost of potentially reduced scientific rigor.
    \item Our approach to identifying relevant keywords was based on a brief, unstructured literature review. This method, while informed by the authors' significant experience and intuitive understanding of commonly used terms in the field, may have inadvertently missed important terms related to systemic risks. Consequently, our search strategy might not have captured all relevant literature.
    \item The process of document screening involved making judgments about whether papers could contain relevant information about AI and systemic risks. Despite our best efforts to maintain objectivity, this subjective element in the screening process may have led to the exclusion of some promising papers that could have contributed valuable insights to the review.
    \item The taxonomy development occurred though conducting a rapid review of the papers. A more comprehensive review, coding approach, and taxonomy development (currently underway) may reveal additional risks and risk sources, and provide further conceptual clarity.
\end{enumerate}

These limitations should be taken into account when considering the comprehensiveness and generalizability of our findings. Future reviews in this rapidly evolving field may benefit from addressing these limitations through expanded database coverage, refined search strategies, and more structured approaches to identifying relevant keywords.

\section{Conclusion}
This systematic review synthesized findings from 86 academic documents examining systemic risks associated with general-purpose AI. Through our rapid review of these papers, we developed an initial taxonomy comprising 13 categories of systemic risks and 50 contributing sources. While this represents a first step toward systematically mapping the landscape of systemic risks from general-purpose AI, further work is needed to refine and validate these classifications through more thorough coding and multiple rounds of taxonomy development.
  
The risks we identified range from operational concerns like control and security, to societal impacts on democracy and fundamental rights, to existential considerations such as irreversible changes to human society. It is important to acknowledge that systemic risks such as these manifest at societal level and do not often have single sources of risk. Furthermore, they are usually not one-off incidents, but rather cumulative and aggregate effects. They frequently have a closer resemblance to gradual phenomena such as climate change or other economic externalities, rather than sudden events. To address these challenges effectively, rigorous and targeted methods for assessing societal-level impacts are needed. 

This descriptive taxonomy represents an initial mapping of how systemic risks and their sources are characterized in the academic literature. While we do not take a position on the plausibility of individual risks or sources, we hope this work provides a foundation for future research and policy development aimed at understanding and addressing the broader societal impacts of general-purpose AI. This future work could enable greater cohesiveness and interoperability across regulations and actors. Integration with existing risk documentation efforts, such as the AI Risk Repository (Slattery et al., 2024), could help ensure decision makers maintain an up-to-date understanding of both risks and their sources.

\section{Acknowledgements}
We thank Annemieke Brouwer, Irene Solaiman, James Gealy, Laura Weidinger, Noam Kolt, and Tekla Emborg for providing very helpful input to this paper. 

\newpage
\bibliography{references}

\newpage

\appendix

\section{Appendices}
\subsection{Excluded documents}
\label{appendix:A}
\begin{longtable}{|p{1cm}|p{7cm}|p{\dimexpr\textwidth-8cm-4\tabcolsep-2\arrayrulewidth}|}
\caption{Reasons for excluding documents} \\
\hline 
\textbf{No} & \textbf{Title} & \textbf{Reason for exclusion} \\
\hline
1 & AI governance and the policymaking process: Key considerations for reducing AI risk & Did not discuss risks \\
\hline
2 & AI, on the Law of the Elephant: Toward Understanding Artificial Intelligence & Did not discuss risks \\
\hline
3 & Artificial intelligence and policy: quo vadis? & Did not discuss risks \\
\hline
4 & Birth of Industry 5.0: Making Sense of Big Data with Artificial Intelligence, "the Internet of Things" and Next-Generation Technology Policy & Did not discuss risks \\
\hline
5 & Calibrating machine behavior: a challenge for AI alignment & Did not discuss risks \\
\hline
6 & Communicating and perceiving risks of artificial intelligence as an emerging technology & Did not find a full text \\
\hline
7 & Defining the Scope of AI Regulations & Did not discuss risks \\
\hline
8 & EVERYDAY AUTOMATION: Experiencing and Anticipating Emerging Technologies & Did not focus on AI \\
\hline
9 & Existential risk from AI and orthogonality: Can we have it both ways? & Did not discuss risks \\
\hline
10 & Global policymakers and catastrophic risk & Did not describe risks \\
\hline
11 & Governing complexity: Design principles for the governance of complex global catastrophic risks & Did not focus on AI \\
\hline
12 & If you worry about humanity, you should be more scared of humans than of AI & Did not discuss large-scale risks \\
\hline
13 & Multiparty Dynamics and Failure Modes for Machine Learning and Artificial Intelligence & Did not discuss risks \\
\hline
14 & Peacekeeping Conditions for an Artificial Intelligence Society & Did not discuss risks \\
\hline
15 & Regulating Artificial Intelligence Proposal for a Global Solution & Did not discuss risks \\
\hline
16 & Responses to the Journey to the Singularity & Did not discuss risks \\
\hline
17 & Should Artificial Intelligence Governance be Centralised? Design Lessons from History & Did not discuss risks \\
\hline
18 & Social choice ethics in artificial intelligence & Did not discuss risks \\
\hline
19 & Social, Organizational, and Individual Impacts of Automation & Did not focus on AI \\
\hline
20 & Space War and AI & Did not discuss risks \\
\hline
21 & The Coming ChatGPT & Did not discuss risks \\
\hline
22 & The emotional impact of generative AI: negative emotions and perception of threat & Did not discuss large-scale risks \\
\hline
23 & Towards an Equitable Digital Society: Artificial Intelligence (AI) and Corporate Digital Responsibility (CDR) & Did not discuss large-scale risks \\
\hline
24 & When Brain Computer Interfaces Pose an Existential Risk & Did not discuss large-scale risks \\
\hline
25 & Generative AI and the Financial Services Industry - Risks according to the International Monetary Fund & Was not academic \\
\hline
26 & Societal Impacts of Artificial Intelligence and Machine Learning & Did not find a full text \\
\hline
\end{longtable}

\subsection{Included documents}
\label{appendix:B}
\begin{longtable}{|p{1cm}|p{7cm}|p{4cm}|p{\dimexpr\textwidth-13cm-4\tabcolsep-2\arrayrulewidth}|}
\caption{Included documents in the review} \\
\hline
\textbf{No} & \textbf{Title} & \textbf{Authors} & \textbf{Year} \\
1  & A reality check and a way forward for the global governance of artificial intelligence.                                                                                                  & Sepasspour, R.                                                                                                                                                                                                                               & 2023 \\ \hline
2  & A Secure Deepfake Mitigation Framework: Architecture, Issues, Challenges, and Societal Impact.                                                                                           & Wazid, M., Mishra, A. K., Mohd, N., Das, A. K.                                                                                                                                                                                               & 2024 \\ \hline
3  & A Survey of the Potential Long-term Impacts of AI: How AI Could Lead to Long-term Changes in Science, Cooperation, Power, Epistemics and Values.                                         & Clarke, S., \& Whittlestone, J.                                                                                                                                                                                                              & 2022 \\ \hline
4  & AI and Catastrophic Risk                                                                                                                                                                 & Bengio, Y.                                                                                                                                                                                                                                   & 2023 \\ \hline
5  & AI and the capitalist space economy.                                                                                                                                                     & Alvarez Leon, L. F.                                                                                                                                                                                                                          & 2021 \\ \hline
6  & AI and the falling sky: interrogating X-Risk.                                                                                                                                            & Jecker, N. S., Atuire, C. A., Bélisle-Pipon, J. C., Ravitsky, V., \& Ho, A.                                                                                                                                                                  & 2024 \\ \hline
7  & AI and Us: Existential Risk or Transformational Tool?                                                                                                                                    & Saavedra-Rivano, N.                                                                                                                                                                                                                          & 2019 \\ \hline
8  & AI-enabled future crime                                                                                                                                                                  & Caldwell, M., Andrews, J.T.A., Tanay, T. et al.                                                                                                                                                                                              & 2020 \\ \hline
9  & An AI Race for Strategic Advantage: Rhetoric and Risks                                                                                                                                   & Cave, S., Ó hÉigeartaigh, S.                                                                                                                                                                                                                 & 2017 \\ \hline
10 & An Interdisciplinary Review of Systemic Risk Factors Leading up to Existential Risks                                                                                                     & Undheim, T. R.                                                                                                                                                                                                                               & 2024 \\ \hline
11 & Arms Race of Artificial Intelligence: Need for International Regulating Mechanism                                                                                                        & Ahmad, K.                                                                                                                                                                                                                                    & 2023 \\ \hline
12 & Artificial Canaries: Early Warning Signs for Anticipatory and Democratic Governance of AI                                                                                                & Zoe Cremer, C., Whittlestone, J.                                                                                                                                                                                                             & 2021 \\ \hline
13 & Artificial Financial Intelligence                                                                                                                                                        & Magnuson, W.                                                                                                                                                                                                                                 & 2020 \\ \hline
14 & Artificial Intelligence and State Economic Security                                                                                                                                      & Uzun, M.                                                                                                                                                                                                                                     & 2020 \\ \hline
15 & Artificial Intelligence and Systemic Risk                                                                                                                                                & Danielsson, J., Macrae, R., Uthemann, A.                                                                                                                                                                                                     & 2019 \\ \hline
16 & Artificial Intelligence and the Ethics of Self-Learning Robots                                                                                                                           & Vallor, S., Bekey, G.A.                                                                                                                                                                                                                      & 2017 \\ \hline
17 & Artificial intelligence challenges in the face of biological threats: emerging catastrophic risks for public health                                                                      & de Lima, R. C., Sinclair, L., Megger, R., Maciel, M. A. G., Vasconcelos, P. F. D. C., \& Quaresma, J. A. S.                                                                                                                                  & 2024 \\ \hline
18 & Artificial Intelligence, Finance, and the Law                                                                                                                                            & Lin, T.C.W.                                                                                                                                                                                                                                  & 2019 \\ \hline
19 & Artificial intelligence, systemic risks, and sustainability                                                                                                                              & Galaz, V., Centeno, M.A., Callahan, P.W., Causevic, A., Patterson, T., Brass, I., Baum, S., Farber, D., Fischer, J., Garcia, D., McPhearson, T., Jimenez, D., King, B., Larcey, P., Levy, K. & 2021 \\ \hline
20 & Artificial Intelligence: Arguments for Catastrophic Risk                                                                                                                                 & Bales, A., D’Alessandro, W., Kirk-Giannini, C. D.                                                                                                                                                                                            & 2024 \\ \hline
21 & Assessing the future plausibility of catastrophically dangerous AI                                                                                                                       & Turchin, A.                                                                                                                                                                                                                                  & 2019 \\ \hline
22 & Autonomous Artificial Intelligence and Uncontemplated Hazards: Towards the Optimal Regulatory Framework                                                                                  & Kovac, M.                                                                                                                                                                                                                                    & 2021 \\ \hline
23 & Beyond the Individual: Governing AI’s Societal Harm                                                                                                                                      & Smuha, N. A.                                                                                                                                                                                                                                 & 2021 \\ \hline
24 & Classification of global catastrophic risks connected with artificial intelligence                                                                                                       & Turchin, A., Denkenberger, D.                                                                                                                                                                                                                & 2020 \\ \hline
25 & Collective action on artificial intelligence: A primer and review                                                                                                                        & de Neufville, R., Baum, S. D.                                                                                                                                                                                                                & 2021 \\ \hline
26 & Comparing minds and machines: implications for financial stability                                                                                                                       & Buckmann, M., Haldane, A., \& Hüser, A. C.                                                                                                                                                                                                   & 2021 \\ \hline
27 & Current and Near-Term AI as a Potential Existential Risk Factor                                                                                                                          & Bucknall, B. S., \& Dori-Hacohen, S.                                                                                                                                                                                                         & 2022 \\ \hline
28 & Cyber factors of strategic stability: How the advance of ai can change the global balance of power                                                                                       & Karasev, P.A.                                                                                                                                                                                                                                & 2020 \\ \hline
29 & Deepfakes and the Weaponization of Disinformation                                                                                                                                        & Brown, N. I.                                                                                                                                                                                                                                 & 2020 \\ \hline
30 & Deepfakes: Trick or treat?                                                                                                                                                               & Kietzmann, J., Lee, L. W., McCarthy, I. P., \& Kietzmann, T. C.                                                                                                                                                                              & 2020 \\ \hline
31 & Defining risk in AI legislation: perspectives and implications in the Korean context                                                                                                     & Kim, J.                                                                                                                                                                                                                                      & 2024 \\ \hline
32 & Designing AI futures: A symbiotic vision                                                                                                                                                 & Gill, K. S.                                                                                                                                                                                                                                  & 2019 \\ \hline
33 & Deterring Algorithmic Manipulation                                                                                                                                                       & Fletcher, G. G. S.                                                                                                                                                                                                                           & 2021 \\ \hline
34 & Developing safer AI-concepts from economics to the rescue                                                                                                                                & Maskara, P. K.                                                                                                                                                                                                                               & 2023 \\ \hline
35 & Discourse analysis of academic debate of ethics for AGI                                                                                                                                  & Graham, R.                                                                                                                                                                                                                                   & 2022 \\ \hline
36 & Emerging Technologies, Risk, Peace, and Conflict                                                                                                                                         & Taylor, N. B.                                                                                                                                                                                                                                & 2023 \\ \hline
37 & Ethics of artificial intelligence                                                                                                                                                        & Liao, S. M.                                                                                                                                                                                                                                  & 2020 \\ \hline
38 & Existential Risks to Humanity Should Concern International Policymakers and More Could Be Done in Considering Them at the International Governance Level                                 & Boyd, M., \& Wilson, N.                                                                                                                                                                                                                      & 2020 \\ \hline
39 & Existential Terrorism: Can Terrorists Destroy Humanity?                                                                                                                                  & Kallenborn, Z., \& Ackerman, G.                                                                                                                                                                                                              & 2023 \\ \hline
40 & Fairness and Bias in Artificial Intelligence: A Brief Survey of Sources, Impacts, and Mitigation Strategies                                                                              & Ferrara, E.                                                                                                                                                                                                                                  & 2023 \\ \hline
41 & Fragmentation and the Future: Investigating Architectures for International AI Governance                                                                                                & Cihon, P., Maas, M. M., \& Kemp, L.                                                                                                                                                                                                          & 2020 \\ \hline
42 & Global health in the age of AI: Safeguarding humanity through collaboration and action                                                                                                   & Guzmán, C. A. F.                                                                                                                                                                                                                             & 2024 \\ \hline
43 & Governing Boring Apocalypses: A new typology of existential vulnerabilities and exposures for existential risk research                                                                  & Liu, H. Y., Lauta, K. C., \& Maas, M. M.                                                                                                                                                                                                     & 2018 \\ \hline
44 & How does artificial intelligence pose an existential risk?                                                                                                                               & Vold, K., \& Harris, D. R.                                                                                                                                                                                                                   & 2021 \\ \hline
45 & International Law Does Not Compute: Artificial Intelligence and the Development, Displacement or Destruction of the Global Legal Order Special Focus: Intersection of Law and Technology & Maas, M. M.                                                                                                                                                                                                                                  & 2019 \\ \hline
46 & Is It a Platform? Is It a Search Engine? It's ChatGPT! The European Liability Regime for Large Language Models Symposium: Artificial Intelligence and Speech                             & Arcila, B. B.                                                                                                                                                                                                                                & 2023 \\ \hline
47 & LLM potentiality and awareness: a position paper from the perspective of trustworthy and responsible AI modeling                                                                         & Sarker, I. H.                                                                                                                                                                                                                                & 2024 \\ \hline
48 & Long-Term Strategies for Ending Existential Risk from Fast Takeoff                                                                                                                       & Dewey, D                                                                                                                                                                                                                                     & 2015 \\ \hline
49 & Managing the ethical and risk implications of rapid advances in artificial intelligence: A literature review                                                                             & Meek, T., Barham, H., Beltaif, N., Kaadoor, A., \& Akhter, T.                                                                                                                                                                                & 2016 \\ \hline
50 & Mankind at a Crossroads: The Future of Our Relation With AI Entities                                                                                                                     & Saavedra-Rivano, N.                                                                                                                                                                                                                          & 2020 \\ \hline
51 & Military artificial intelligence as a contributor to global catastrophic risk                                                                                                            & Maas, M. M., Lucero-Matteucci, K., \& Cooke, D.                                                                                                                                                                                              & 2023 \\ \hline
52 & Moral circle expansion: A promising strategy to impact the far future                                                                                                                    & Anthis, J. R., \& Paez, E.                                                                                                                                                                                                                   & 2021 \\ \hline
53 & Motivations for Artificial Intelligence, for Deep Learning, for ALife: Mortality and Existential Risk                                                                                    & Harvey, I.                                                                                                                                                                                                                                   & 2024 \\ \hline
54 & Nudging Robots: Innovative Solutions to Regulate Artificial Intelligence                                                                                                                 & Guihot, M., Matthew, A. F., \& Suzor, N. P.                                                                                                                                                                                                  & 2017 \\ \hline
55 & Optimising peace through a Universal Global Peace Treaty to constrain the risk of war from a militarised artificial superintelligence                                                    & Carayannis, E.G., Draper, J.                                                                                                                                                                                                                 & 2023 \\ \hline
56 & Pause artificial intelligence research? Understanding AI policy challenges                                                                                                               & Goldfarb, A.                                                                                                                                                                                                                                 & 2024 \\ \hline
57 & Possible harms of artificial intelligence and the EU AI act: fundamental rights and risk                                                                                                 & Kusche, I.                                                                                                                                                                                                                                   & 2024 \\ \hline
58 & Regulate against the machine: how the EU mitigates AI harm to democracy                                                                                                                  & Cupać, J., \& Sienknecht, M.                                                                                                                                                                                                                 & 2024 \\ \hline
59 & Regulating Artificial Intelligence Systems: Risks, Challenges, Competencies, and Strategies                                                                                              & Scherer, M. U.                                                                                                                                                                                                                               & 2016 \\ \hline
60 & Regulating for `Normal Al Accidents': Operational Lessons for the Responsible Governance of Artificial Intelligence Deployment                                                           & Maas, M. M.                                                                                                                                                                                                                                  & 2018 \\ \hline
61 & Regulating high-reach AI: On transparency directions in the Digital Services Act                                                                                                         & Söderlund, K., Engström, E., Haresamudram, K., Larsson, S., \& Strimling, P.                                                                                                                                                                 & 2024 \\ \hline
62 & Regulation Priorities for Artificial Intelligence Foundation Models                                                                                                                      & Gaske, M. R.                                                                                                                                                                                                                                 & 2023 \\ \hline
63 & Responses to catastrophic AGI risk: a survey                                                                                                                                             & Sotala, K., \& Yampolskiy, R. V.                                                                                                                                                                                                             & 2014 \\ \hline
64 & Responsible artificial intelligence in agriculture requires systemic understanding of risks and externalities                                                                            & Tzachor, A., Devare, M., King, B., Avin, S., \& Ó hÉigeartaigh, S.                                                                                                                                                                           & 2022 \\ \hline
65 & Rethinking AI Strategy and Policy as Entangled Super Wicked Problems                                                                                                                     & Gruetzemacher, R.                                                                                                                                                                                                                            & 2018 \\ \hline
66 & Risks of the Journey to the Singularity                                                                                                                                                  & Sotala, K., \& Yampolskiy, R.                                                                                                                                                                                                                & 2017 \\ \hline
67 & Safely advancing a spacefaring humanity with artificial intelligence                                                                                                                     & Richards, C. E., Cernev, T., Tzachor, A., Zilgalvis, G., \& Kaleagasi, B.                                                                                                                                                                    & 2023 \\ \hline
68 & Singularity and Coordination Problems: Pandemic Lessons from 2020                                                                                                                        & Corrêa, N. K., \& De Oliveira, N.                                                                                                                                                                                                            & 2021 \\ \hline
69 & Societal and ethical impacts of artificial intelligence: Critical notes on European policy frameworks                                                                                    & Vesnic-Alujevic, L., Nascimento, S., \& Polvora, A.                                                                                                                                                                                          & 2020 \\ \hline
70 & Societal impacts of chatbot and mitigation strategies for negative impacts: A large-scale qualitative survey of ChatGPT users                                                            & Wei, X., Chu, X., Geng, J., Wang, Y., Wang, P., Wang, H., ... \& Lei, L.                                                                                                                                                                     & 2024 \\ \hline
71 & Sociotechnical Harms of Algorithmic Systems: Scoping a Taxonomy for Harm Reduction                                                                                                       & Shelby, R., Rismani, S., Henne, K., Moon, A., Rostamzadeh, N., Nicholas, P., ... \& Virk, G.                                                                                                                                                 & 2024 \\ \hline
72 & The argument for near-term human disempowerment through AI                                                                                                                               & Dung, L.                                                                                                                                                                                                                                     & 2024 \\ \hline
73 & The blended future of automation and AI: Examining some long-term societal and ethical impact features                                                                                   & Khogali, H. O., \& Mekid, S.                                                                                                                                                                                                                 & 2023 \\ \hline
74 & The case for a broader approach to AI assurance: addressing “hidden” harms in the development of artificial intelligence                                                                 & Thomas, C., Roberts, H., Mökander, J. et al.                                                                                                                                                                                                 & 2024 \\ \hline
75 & The European Union’s Approach to Artificial Intelligence and the Challenge of Financial Systemic Risk                                                                                    & Keller, A., Pereira, C. M., \& Pires, M. L.                                                                                                                                                                                                  & 2024 \\ \hline
76 & The global catastrophic risks... ...Connected with the possibility of finding alien Al during SETI                                                                                       & Turchin, A.                                                                                                                                                                                                                                  & 2018 \\ \hline
77 & The possibility and risks of artificial general intelligence                                                                                                                             & Torres, P.                                                                                                                                                                                                                                   & 2019 \\ \hline
78 & The Power Structure of Artificial Intelligence                                                                                                                                           & Liu, H. Y.                                                                                                                                                                                                                                   & 2018 \\ \hline
79 & The risks associated with Artificial General Intelligence: A systematic review                                                                                                           & McLean, S., Read, G. J., Thompson, J., Baber, C., Stanton, N. A., \& Salmon, P. M.                                                                                                                                                           & 2023 \\ \hline
80 & The Risks of Low Level Narrow Artificial Intelligence                                                                                                                                    & Page, J., Bain, M., \& Mukhlish, F.                                                                                                                                                                                                          & 2018 \\ \hline
81 & The Societal Impacts of Generative Artificial Intelligence: A Balanced Perspective                                                                                                       & Sabherwal, R., \& Grover, V                                                                                                                                                                                                                  & 2024 \\ \hline
82 & The transformative potential of artificial intelligence                                                                                                                                  & Gruetzemacher, R., \& Whittlestone, J.                                                                                                                                                                                                       & 2022 \\ \hline
83 & The united nation's capability to manage existential risks with a focus on artificial intelligence                                                                                       & Nindler, R.                                                                                                                                                                                                                                  & 2019 \\ \hline
84 & Understanding Artificial Agency                                                                                                                                                          & Dung, L.                                                                                                                                                                                                                                     & 2024 \\ \hline
85 & What Is an Artificial Intelligence Arms Race Anyway                                                                                                                                      & Asaro, P.                                                                                                                                                                                                                                    & 2019 \\ \hline
86 & Why we are failing to understand the societal impact of artificial intelli                                                                                                               & Jaume-Palasi, L.                                                                                                                                                                                                                             & 2019 \\ 
\hline
\end{longtable}

\subsection{Types of systemic risks}
\begin{longtable}{|p{7cm}|p{\dimexpr\textwidth-7cm-4\tabcolsep-2\arrayrulewidth}|}
\caption{Select quotes about types of systemic risks} \\
\hline
\textbf{Quote} & \textbf{Source} \\
\hline
Within that multiplicity of AI systems, a variety of forms of self-awareness will emerge, and it isconceivable and even likely that at least some of them will be hostile to us having control over them                                                                                                                                                                                                                                                                                                                                                                                                                                                                                                                                                                                                                                                                                                                                                                                                                                                                                                                                                                                                                                                                                                                                                                                                                                                                                         & Mankind at a Crossroads: The Future of Our Relation With AI Entities                         \\ \hline
If successful, these technologies would result in the “design” of super-intelligent humans, thatis, humans with intellectual capabilities matching those of AI systems. Those people would be ableto communicate with machines at their level and, presumably, control them despite them having achieved sentience.                                                                                                                                                                                                                                                                                                                                                                                                                                                                                                                                                                                                                                                                                                                                                                                                                                                                                                                                                                                                                                                                                                                                                                              & Mankind at a Crossroads: The Future of Our Relation With AI Entities                         \\ \hline
emergence of a “brave newworld” where an elite of super humans would control and dominate an underclass of normal people.                                                                                                                                                                                                                                                                                                                                                                                                                                                                                                                                                                                                                                                                                                                                                                                                                                                                                                                                                                                                                                                                                                                                                                                                                                                                                                                                                                        & Mankind at a Crossroads: The Future of Our Relation With AI Entities                         \\ \hline
To be able to disempower humanity, an AI would likely need to be able to form and execute long-term plans. To the extent that the system cannot execute on long-term plans, there is low risk that it can accumulate power to achieve some other goal                                                                                                                                                                                                                                                                                                                                                                                                                                                                                                                                                                                                                                                                                                                                                                                                                                                                                                                                                                                                                                                                                                                                                                                                                                            & Understanding Artificial Intelligence                                                        \\ \hline
loss of control of an AIsystem—typically when it is given or develops a strong self-preserva-tion goal, possibly creating an existential threat to humanity—which canhappen intentionally or not, and illustrates a misalignment between theAI and both the human operator and society.                                                                                                                                                                                                                                                                                                                                                                                                                                                                                                                                                                                                                                                                                                                                                                                                                                                                                                                                                                                                                                                                                                                                                                                                          & AI and Catastrophic Risk                                                                     \\ \hline
Unfortunately,even slight misalignment can result in the emergence of unintendedgoals, such as power-seeking and self-preservation, which can be ad-vantageous for accomplishing virtually any other goal. For instance, ifan AI acquires a self-preservation goal, it will resist attempts to shut itdown, creating immediate conflict and potentially leading to a loss ofcontrol if we fail to deactivate it.                                                                                                                                                                                                                                                                                                                                                                                                                                                                                                                                                                                                                                                                                                                                                                                                                                                                                                                                                                                                                                                                                 & AI and Catastrophic Risk                                                                     \\ \hline
Tails riskspertain to the probability (Bostrom 2002) of losing controlover not just the segment but “loss of control of our civili-zation” (Schechner 2023) to AI                                                                                                                                                                                                                                                                                                                                                                                                                                                                                                                                                                                                                                                                                                                                                                                                                                                                                                                                                                                                                                                                                                                                                                                                                                                                                                                                & Developing safer AI–concepts from economics to the rescue                                    \\ \hline
This ‘gorilla problem’, as Russell terms it, is the idea behind P2, which states that ifan AI system has a decisive strategic advantage over human intelligence, then we may not be able to controlthat system.                                                                                                                                                                                                                                                                                                                                                                                                                                                                                                                                                                                                                                                                                                                                                                                                                                                                                                                                                                                                                                                                                                                                                                                                                                                                                  & How Does Articial Intelligence Pose an Existential Risk?                                     \\ \hline
Early concerns around AI Xrisks focused on the possibility of an intelligenceexplosion and the subsequent pathway to a scenario in which a powerful superintelligent AI has misalignedobjectives from humanity. These concerns have not gone away, but they have evolved over time                                                                                                                                                                                                                                                                                                                                                                                                                                                                                                                                                                                                                                                                                                                                                                                                                                                                                                                                                                                                                                                                                                                                                                                                               & How Does Articial Intelligence Pose an Existential Risk?                                     \\ \hline
nambiguously,there are short-term risks from misinformation and job loss and real longer-term worriesabout inequality, democracy and contro                                                                                                                                                                                                                                                                                                                                                                                                                                                                                                                                                                                                                                                                                                                                                                                                                                                                                                                                                                                                                                                                                                                                                                                                                                                                                                                                                      & Pause artiﬁcial intelligence research? Understanding AI policy challenges                    \\ \hline
The risks created by the autonomy of Al encompass not onlyproblems of foreseeability, but also problems of control. It might bedifficult for humans to maintain control of machines that are pro-grammed to act with considerable autonomy. There are any number ofmechanisms by which a loss of control may occur: a malfunction,such as a corrupted file or physical damage to input equipment; a se-curity breach; the superior response time of computers as compared tohumans;5 ' or flawed programming. The last possibility raises the mostinteresting issues because it creates the possibility that a loss of con-trol might be the direct but unintended consequence of a consciousdesign choice. Control, once lost, may be difficult to regain if the Alis designed with features that permit it to learn and adapt. These arethe characteristics that make Al a potential source of public risk on a scale that far exceeds the more familiar forms of public risk that aresolely the result of human behavior. scale that far exceeds the more familiar forms of public risk that aresolely the result of human behavior. Loss of control can be broken down into two varieties. A loss oflocal control occurs when the Al system can no longer be controlledby the human or humans legally responsible for its operation and su-pervision. A loss of general control occurs when the Al system can nolonger be controlled by any human. & Regulating Artificial Intelligence Systems                                                    \\ \hline
The risks associated with containment, confinement, andcontrol in the AGI development phase, and after an AGI hasbeen developed, loss of control of an AGI                                                                                                                                                                                                                                                                                                                                                                                                                                                                                                                                                                                                                                                                                                                                                                                                                                                                                                                                                                                                                                                                                                                                                                                                                                                                                                                                       & The risks associated with Artificial General Intelligence: A systematic review               \\ \hline
These suggestions are reasonable in principle, but it's unclear that they'll be easy to implement in practice.Ensuring that AI systems do what we want requires ruling out misaligned behavior contrary to designers' in-tentions.5 If misaligned behavior proves difficult to foresee prior to deployment, as some suspect it will, then well‐meaning developers and regulators may lack the tools to avert catastrophe.                                                                                                                                                                                                                                                                                                                                                                                                                                                                                                                                                                                                                                                                                                                                                                                                                                                                                                                                                                                                                                                                        & Artificial Intelligence: Arguments for Catastrophic Risk                                     \\ \hline
Dialectically, the Singularity Hypothesis makes contact with the Problem of Power‐Seeking at two points: first,misaligned agential AI systems with instrumentally convergent reasons to engage in self‐improvement might triggera singularity‐type recursive process; second, human actors seeking to improve the capabilities of AI systems couldinitiate a singularity‐type recursive process which gives rise to misaligned supercapable AI systems                                                                                                                                                                                                                                                                                                                                                                                                                                                                                                                                                                                                                                                                                                                                                                                                                                                                                                                                                                                                                                           & Artificial Intelligence: Arguments for Catastrophic Risk                                     \\ \hline
The failure to do so effectively and swiftly might put us in a race to weaponize AI or haveAI escape our control (i.e., misalign with our own goals) with potentially catastrophic conse-quences                                                                                                                                                                                                                                                                                                                                                                                                                                                                                                                                                                                                                                                                                                                                                                                                                                                                                                                                                                                                                                                                                                                                                                                                                                                                                                 & Global health in the age of AI: Safeguardinghumanity through collaboration and action        \\ \hline
The potential for the misalignment of interests flows from the factthat an Al's objectives are determined by its initial programming... In such a scenario, the risk from an Al system does not stem frommalevolence or an inability to comprehend the subjective intent be-hind its programmed goals. Rather, it stems from the machine's fun-damental indifference to that subjective intent. A growing chorus of academics, tech entrepreneurs, and futuristshas gone further, warning that stronger forms of Al may resist all hu-man efforts to govern their actions and pose a catastrophic - perhapseven existential - risk to humanity.                                                                                                                                                                                                                                                                                                                                                                                                                                                                                                                                                                                                                                                                                                                                                                                                                                                  &                                                                                              \\ \hline
first, the speed of advances in AI capability, as well as the capability level current systems have already reached, suggestthat it is practically possible to build AI systems capable of disempowering humanity by 2100. Second, due to incentivesand coordination problems, if it is possible to build such AI, it will be built. Third, since it appears to be a hard technicalproblem to build AI which is aligned with the goals of its designers, and many actors might build powerful AI, misalignedpowerful AI will be built. Fourth, because disempowering humanity is useful for a large range of misaligned goals, such AIwill try to disempower humanity. If AI is capable of disempowering humanity and tries to disempower humanity by 2100,then humanity will be disempowered by 2100.                                                                                                                                                                                                                                                                                                                                                                                                                                                                                                                                                                                                                                                                                           & The argument for near‑term human disempowerment through AI                                   \\ \hline
The existential threats perspective provides a third insight into AI power.77 Thislens foregrounds an under-examined aspect of power: that of uniform appli-cation over humanity as a whole. Power discourses predominantly focusupon relational dynamics and disparate impacts, whereby individuals orgroups wield power over others on a sliding scale. The prospect of AI powerwithin the frame of existential threats is so overwhelming that it will eliminatethe disparities between individuals in the experience of that power either bytriggering human extinction or curbing the future of humanity.                                                                                                                                                                                                                                                                                                                                                                                                                                                                                                                                                                                                                                                                                                                                                                                                                                                                                   & The power structure of artiﬁcial intelligence                                                \\ \hline
In conclusion, algorithmic systems and artificial intelligenceas collectivistic technologies amplify a weakness of democracies:the methodological individualism of democratic systems has char-acterized the normative approach of democratic powers.                                                                                                                                                                                                                                                                                                                                                                                                                                                                                                                                                                                                                                                                                                                                                                                                                                                                                                                                                                                                                                                                                                                                                                                                                                            & Why We Are Failing to Understand the Societal Impact of Artificial Intelligence              \\ \hline
The introduction ofdeepfakes technology causes a signiﬁcant problem. As it can modifycomments, speeches, or events involving inﬂuential politicians. Dueto this, it introduces a potential threat to the political system of ademocracy. The distribution of those fabricated materials producesa great inﬂuence on public sentiments. It may cause social distur-bances and can aﬀect election results                                                                                                                                                                                                                                                                                                                                                                                                                                                                                                                                                                                                                                                                                                                                                                                                                                                                                                                                                                                                                                                                                          & A Secure Deepfake Mitigation Framework: Architecture, Issues, Challenges,and Societal Impact \\ \hline
Developments in AI are giving companies and governmentsmore control over individuals’ lives than ever before, and maypossibly be used to undermine democratic processes                                                                                                                                                                                                                                                                                                                                                                                                                                                                                                                                                                                                                                                                                                                                                                                                                                                                                                                                                                                                                                                                                                                                                                                                                                                                                                                          & A Survey of the Potential Long-term Impacts of AI                                            \\ \hline
In the extreme,a few individuals controlling superhuman AIs would accrue a level ofpower never before seen in human history, a blatant contradiction withthe very principle of democracy and a major threat to it                                                                                                                                                                                                                                                                                                                                                                                                                                                                                                                                                                                                                                                                                                                                                                                                                                                                                                                                                                                                                                                                                                                                                                                                                                                                                & AI and catastrophic risk                                                                     \\ \hline
A fourth risk is AI’s capacity to turbocharge misinformation bymeans of LLMs and deep fakes in ways that undermine autonomyand democracy                                                                                                                                                                                                                                                                                                                                                                                                                                                                                                                                                                                                                                                                                                                                                                                                                                                                                                                                                                                                                                                                                                                                                                                                                                                                                                                                                         & AI and the falling sky: interrogating X-Risk                                                 \\ \hline
Misinformation {[}72{]} and targeted messaging {[}73{]} can havetransformative implications for the resilience of democracies and verypossibility of collective action {[}74{]}, {[}75{]}                                                                                                                                                                                                                                                                                                                                                                                                                                                                                                                                                                                                                                                                                                                                                                                                                                                                                                                                                                                                                                                                                                                                                                                                                                                                                                        & Artificial Canaries: Early Warning Signs forAnticipatory and Democratic Governance of AI     \\ \hline
The fact that certain uses of AI-systems for instance risk harming the democ-ratic process, eroding the rule of law or exacerbating inequality goes beyond theconcern of (the sum of) individuals but affects society at large                                                                                                                                                                                                                                                                                                                                                                                                                                                                                                                                                                                                                                                                                                                                                                                                                                                                                                                                                                                                                                                                                                                                                                                                                                                                   & Beyond the individual: governing AI’ssocietal harm                                           \\ \hline
The three examples referred to above—AI-based facial recognition, AI-based votermanipulation, and AI-based public decision-making—concern three different AI ap-plications impacting (at least) three different societal interests: equality, democra-cy and the rule of law.                                                                                                                                                                                                                                                                                                                                                                                                                                                                                                                                                                                                                                                                                                                                                                                                                                                                                                                                                                                                                                                                                                                                                                                                                    & Beyond the individual: governing AI’ssocietal harm                                           \\ \hline
Many of the AI trends and impacts discussed above have directimplications on the functioning of social and political structures.For example, the use of recommender systems and incentives forbrevity on social media may lead to an increase in political polarisa-tion and a less informed public respectively. In democratic societiesthis may have a knock-on effect in terms of the political process. Anincrease in political polarisation in the general public, for example,may be reflected within political institutions, whereby elected rep-resentatives are unwilling to compromise due to a rise in politicalpartisanship. This would lead to an increase in difficulty of passingkey legislature aimed at addressing pressing problems, includingexistential risks, thereby acting as a risk factor. Furthermore, aless informed public may be more susceptible to coordinated mis-and dis-information campaigns, such as the Russian state interfer-ence with the 2016 U.S. presidential election {[} 61 {]}. Such effects actas risk factors by eroding trust in traditional democratic politicalstructures and processes, thus inhibiting their ability to respondto existential risks. Political processes may also be hindered bychanges to state-corporation relations that result in corporationshaving greater political lobbying power, or political division overgovernment regulation of multinational corporations                                                    & Current and Near-Term AI as a Potential Existential Risk Factor                              \\ \hline
These harms are not speculative. Deepfake nonconsensualpornography has already appeared online, and the United States hasexperienced foreign interference with our democratic discourse andelections. And the threat of further disruption is compounded by theunchecked power of social media.                                                                                                                                                                                                                                                                                                                                                                                                                                                                                                                                                                                                                                                                                                                                                                                                                                                                                                                                                                                                                                                                                                                                                                                                  & Deepfakes and the weaponization of Disinformation                                            \\ \hline
Many deepfakes will be publicly distributed,particularly those aimed at disrupting democratic discourse,undermining diplomacy, exacerbating social divisions,threatening public safety and national security, exactingrevenge through the distribution of nonconsensualpornography, and otherwise inflicting widespread harms.                                                                                                                                                                                                                                                                                                                                                                                                                                                                                                                                                                                                                                                                                                                                                                                                                                                                                                                                                                                                                                                                                                                                                                   & Deepfakes and the weaponization of Disinformation                                            \\ \hline
Deepfakepropaganda and election meddling, and the disin-formation they seed threaten efﬁcient governancefor all democracies if not democracy itself                                                                                                                                                                                                                                                                                                                                                                                                                                                                                                                                                                                                                                                                                                                                                                                                                                                                                                                                                                                                                                                                                                                                                                                                                                                                                                                                              & Deepfakes: Trick or Treat?                                                                   \\ \hline
From a boarder societal perspective, the future of AI poses challenges of democraticpolitics, including questions of political agency, accountability, and representation                                                                                                                                                                                                                                                                                                                                                                                                                                                                                                                                                                                                                                                                                                                                                                                                                                                                                                                                                                                                                                                                                                                                                                                                                                                                                                                        & Designing AI Futures: A Symbiotic Vision                                                     \\ \hline
technology is developing so quickly that regulators simply cannot catch up: therhythm of market-driven innovation generally outpaces the democratic legisla-tive process                                                                                                                                                                                                                                                                                                                                                                                                                                                                                                                                                                                                                                                                                                                                                                                                                                                                                                                                                                                                                                                                                                                                                                                                                                                                                                                         & Ethical Principles and Governance for AI                                                     \\ \hline
Think of thecapacity to create deep fakes intended to damage someone’s reputation; or thevery real possibility of manipulating public opinion or even destabilizing demo-cratic processes with biased or false information targeted to the most vulnerableor gullible audience, etc                                                                                                                                                                                                                                                                                                                                                                                                                                                                                                                                                                                                                                                                                                                                                                                                                                                                                                                                                                                                                                                                                                                                                                                                              & Ethical Principles and Governance for AI                                                     \\ \hline
All these com-pounded risks threaten to distort political debate, affect democracy, and even en-danger public safety.9 Additionally, OpenAI reported an estimated 100 million ac-tive users of ChatGPT in January 2023, which makes the potential for a vast andsystemic impact of these risks a considerable one                                                                                                                                                                                                                                                                                                                                                                                                                                                                                                                                                                                                                                                                                                                                                                                                                                                                                                                                                                                                                                                                                                                                                                                & Is It a Platform ? Is It a Search Engine ? It's ChatGPT                                      \\ \hline
nambiguously,there are short-term risks from misinformation and job loss and real longer-term worriesabout inequality, democracy and control                                                                                                                                                                                                                                                                                                                                                                                                                                                                                                                                                                                                                                                                                                                                                                                                                                                                                                                                                                                                                                                                                                                                                                                                                                                                                                                                                     & Pause artiﬁcial intelligence research? Understanding AI policy challenges                    \\ \hline
In recent years AI-powered techniques such as proﬁling, targeting, election manipulation, and massive disinformation campaigns via social bots and troll farms challenge the very foundations of democratic systems                                                                                                                                                                                                                                                                                                                                                                                                                                                                                                                                                                                                                                                                                                                                                                                                                                                                                                                                                                                                                                                                                                                                                                                                                                                                              & Regulate against the machine: how the EU mitigates AIharm to democracy                       \\ \hline
Alternately, mass unemployment from job auto-mation in democratic societies could lead to the election ofpopulist officials4 (Frey, Berger et al. 2017) which couldhave numerous and unanticipated destabilizing effects                                                                                                                                                                                                                                                                                                                                                                                                                                                                                                                                                                                                                                                                                                                                                                                                                                                                                                                                                                                                                                                                                                                                                                                                                                                                         & Rethinking AI Strategy and Policy as Entangled Super Wicked Problems                         \\ \hline
Political and civic harms; Political harms emerge when “peo-ple are disenfranchised and deprived of appropriate political powerand influence” {[}186, p. 162{]}. These harms focus on the domain ofgovernment, and focus on how algorithmic systems govern throughindividualized nudges or micro-directives {[} 187{]}, that may destabi-lize governance systems, erode human rights, be used as weaponsof war {[} 188{]}, and enact surveillant regimes that disproportion-ately target and harm people of color {[} 120{]}. More generally, theseharms may erode democracy {[}97 {]}, through election interferenceor censorship {[}207{]}. Moreover, algorithmic systems may exacer-bate social inequalities and reduction of civil liberties within legalsystems {[}139, 181{]}, such as unreasonable searches {[}152{]}, wrongfularrest {[} 61 , 61 , 124{]}, or court transcription errors {[}124{]}. These harmsadversely impact how a nation’s institutions or services function {[} 3 {]}and increase societal polarization {[}207{]}                                                                                                                                                                                                                                                                                                                                                                                                                                                   & Sociotechnical Harms of Algorithmic Systems: Scoping aTaxonomy for Harm Reduction            \\ \hline
failure in regulatory control couldresult in bad actors spreading more misinformation atscale, which would undermine democracy                                                                                                                                                                                                                                                                                                                                                                                                                                                                                                                                                                                                                                                                                                                                                                                                                                                                                                                                                                                                                                                                                                                                                                                                                                                                                                                                                                   & Societal Impacts of Generative AI: A Balanced Perspective                                    \\ 
\hline
\end{longtable}

\subsection{Sources of systemic risks}
\begin{longtable}{|p{7cm}|p{\dimexpr\textwidth-7cm-4\tabcolsep-2\arrayrulewidth}|}
\caption{Select quotes about sources of systemic risks} \\
\hline
\textbf{Quote} & \textbf{Source} \\
\hline
The correlations drawn by the AI system can be inapplicable, biased or incorrect, but it is also possible that the legal rule applicable to the situation was erroneously em-bedded in the system.                                                                                                                                                            & Beyond the Individual: Governing AI’s Societal Harm                                                                                   \\ \hline
lack of transparency of how AI systems are used by the product or service provider.                                                                                                                                                                                                                                                                           & Beyond the Individual: Governing AI’s Societal Harm                                                                                   \\ \hline
difficult to demonstrate it and establish a causal link. I call this the knowledge gap problem                                                                                                                                                                                                                                                                & Beyond the Individual: Governing AI’s Societal Harm                                                                                   \\ \hline
even if there is awareness, the individual harm may be perceived as insignificant, or in any case as too small in proportion to the costs that may be involved when challenging it. Hence, the individual is unlikely prompted to challenge the problematic practice. I call this the threshold problem.                                                      & Beyond the Individual: Governing AI’s Societal Harm                                                                                   \\ \hline
the current legal framework primarily focuses on legal remedies for the individuals directly subject- ed to the practice, rather than to ‘society’                                                                                                                                                                                                            & Beyond the Individual: Governing AI’s Societal Harm                                                                                   \\ \hline
differently than individual or collective harm, societal harm will often mani-fest itself at a subsequent stage only, namely over the longer term rather than in the immediate period following the AI system’s use                                                                                                                                           & Beyond the Individual: Governing AI’s Societal Harm                                                                                   \\ \hline
his gap in time—especially when combined with the opacity of the AI systems functioning and use—not only complicates the identification of the harm itself, but also of the causal link be- tween the harm and the AI-related practice                                                                                                                        & Beyond the Individual: Governing AI’s Societal Harm                                                                                   \\ \hline
An additional obstacle in this regard concerns the ‘virtual’ nature of the harm. In contrast with, for instance, the harm caused by a massive oil leak or burning forest, the societal harm that can arise from the use of certain AI applications is not as tangible, visible, measurable or predictable                                                     & Beyond the Individual: Governing AI’s Societal Harm                                                                                   \\ \hline
societal harm typically does not arise from a single occurrence of the prob-lematic AI practice. Instead, it is often the widespread, repetitive or accumulative character of the practice that can render it harmful from a societal perspective                                                                                                             & Beyond the Individual: Governing AI’s Societal Harm                                                                                   \\ \hline
challenging to in-stil a sense of responsibility in those who instigate a certain practice, if it is only the accumulative effect of their action together with the action of other actors that causes the harm                                                                                                                                               & Beyond the Individual: Governing AI’s Societal Harm                                                                                   \\ \hline
Besides the accumulative effects raised by a certain type of con-duct, also the opacity of different types of (interacting) conduct by many hands can contribute to the harm and to the difficulty of identifying and addressing it                                                                                                                           & Beyond the Individual: Governing AI’s Societal Harm                                                                                   \\ \hline
at the level of the AI system, multiple compo-nents developed and operated by multiple actors can interact with each other and cause harm, without it being clear which component or interaction is the direct contributor thereof                                                                                                                            & Beyond the Individual: Governing AI’s Societal Harm                                                                                   \\ \hline
at the level of the organisation or institution deploy-ing the system (whether in the public or private sector), different individuals may contribute to a process in many different ways, whereby the resulting practice can cause harm                                                                                                                      & Beyond the Individual: Governing AI’s Societal Harm                                                                                   \\ \hline
this issue manifests itself at the level of the network of organisa-tions and institutions that deploy the problematic AI application. The scale of these networks and their potential interconnectivity and interplay renders the identification of the problematic cause virtually impossible—even more so if there isn’t necessarily one problematic cause & Beyond the Individual: Governing AI’s Societal Harm                                                                                   \\ \hline
the societal harm that can arise from the mentioned AI applications is not easily expressible in current human rights dis- course (Yeung, 2019a). While a particular human right may well be impacted, a one-on-one relationship between such right and the societal harm in question can be lacking                                                          & Beyond the Individual: Governing AI’s Societal Harm                                                                                   \\ \hline
A growing reliance on artificial intelligence and other forms of technology in the financial industry can exacerbate intertwined systemic risks related to size, speed, and interconnectivity. Moreover, the growing complexity of technology increases the risks of serious financial accidents                                                              & Artificial Intelligence, Finance, and the Law                                                                                         \\ \hline
Wider adoption of financial artificial intelligence can amplify certain systemic risks for the financial system relating to size, speed, and linkage                                                                                                                                                                                                          & Artificial Intelligence, Finance, and the Law                                                                                         \\ \hline
sudden discontinuity in which AGIs rapidly become far more numerous or intelligent due to conceptual breakthroughs, fast computing hardware, self-improvement capability                                                                                                                                                                                      & Responses to catastrophic AGI risk: a survey                                                                                          \\ \hline
As machines become more autonomous, humans will have fewer opportunities to intervene in time and will be forced to rely on machines making good choices                                                                                                                                                                                                      & Responses to catastrophic AGI risk: a survey                                                                                          \\ \hline
AGIs were both powerful and indifferent to human values                                                                                                                                                                                                                                                                                                       & Responses to catastrophic AGI risk: a survey                                                                                          \\ \hline
under plausible assumptions, some AI systems will seek power, successfully obtain it and go awry, potentially bringing about catastrophe                                                                                                                                                                                                                      & Artificial Intelligence: Arguments for Catastrophic Risk                                                                              \\ \hline
AIs with radically greater cognitive powers than the smartest humans.                                                                                                                                                                                                                                                                                         & Artificial Intelligence: Arguments for Catastrophic Risk                                                                              \\ \hline
conflict with humans as competition arises for resources and influence                                                                                                                                                                                                                                                                                        & Artificial Intelligence: Arguments for Catastrophic Risk                                                                              \\ \hline
deceptive alignment, which involves systems appearing safe during development but becoming dangerous when they're deployed in the world                                                                                                                                                                                                                       & Artificial Intelligence: Arguments for Catastrophic Risk                                                                              \\ \hline
even if these risks were real and there was mounting evidence for this, some people might not take the danger seriously                                                                                                                                                                                                                                       & Artificial Intelligence: Arguments for Catastrophic Risk                                                                              \\ \hline
reward misspecification                                                                                                                                                                                                                                                                                                                                       & Artificial Intelligence: Arguments for Catastrophic Risk                                                                              \\ \hline
goal misgeneralization                                                                                                                                                                                                                                                                                                                                        & Artificial Intelligence: Arguments for Catastrophic Risk                                                                              \\ \hline
certain resource‐acquiring, self‐improving and shutdown‐resisting subgoals                                                                                                                                                                                                                                                                                    & Artificial Intelligence: Arguments for Catastrophic Risk                                                                              \\ \hline
ability to reflect on and modify one's goals                                                                                                                                                                                                                                                                                                                  & Artificial Intelligence: Arguments for Catastrophic Risk                                                                              \\ \hline
the interplay of technical systems and social factors {[} 35 , 90 {]} and can encode systemic inequalitie, often adopt the default norms, and power structures of society                                                                                                                                                                                     & Sociotechnical Harms of Algorithmic Systems: Scoping a Taxonomy for Harm Reduction                                                    \\ \hline
existing and intersecting power dynamics continued within sociotechnical systems                                                                                                                                                                                                                                                                              & Sociotechnical Harms of Algorithmic Systems: Scoping a Taxonomy for Harm Reduction                                                    \\ \hline
widespread, repetitive or accumulative character of algorithmic systems in the world {[}207, p. 10{]}, which contribute to institutional exclusions                                                                                                                                                                                                           & Sociotechnical Harms of Algorithmic Systems: Scoping a Taxonomy for Harm Reduction                                                    \\ \hline
more capable regarding certain types of tasks than humans                                                                                                                                                                                                                                                                                                     & Singularity and Coordination Problems: Pandemic Lessons from 2020                                                                     \\ \hline
develop goals and objectives that diverge from us humans                                                                                                                                                                                                                                                                                                      & Singularity and Coordination Problems: Pandemic Lessons from 2020                                                                     \\ \hline
data collection and storage, as well as model training require energy intensive computational power                                                                                                                                                                                                                                                           & The case for a broader approach to AI assurance: addressing “hidden” harms in the development of artificial intelligence              \\ \hline
outsourcing of “data labelling”, and Reinforcement Learning from Human Feedback (RLHF) tasks to low income countries                                                                                                                                                                                                                                          & The case for a broader approach to AI assurance: addressing “hidden” harms in the development of artificial intelligence              \\ \hline
deploying AI at scale: the influence of a few common ML platforms, consequently creating centralized points of failure, where deliberate attacks could cause disproportionate harm                                                                                                                                                                            & Responsible artificial intelligence in agriculture requires systemic understanding of risks and externalities                         \\ \hline
humanity’s quest for ASI-enabled warfare & Optimising peace through a Universal Global Peace Treaty to constrain the risk of war from a militarised artificial superintelligence \\
\hline
\end{longtable}

\end{document}